\documentclass[a4paper]{article}

\title{Global-Local View: Scalable Consistency for Concurrent Data Types} 
\usepackage{authblk}

\author[1]{Deepthi Devaki Akkoorath}
\author[3]{Jos\'e Brand\~ao}
\author[2]{Annette Bieniusa}
\author[4]{Carlos Baquero}
\affil[1]{Technical University of Kaiserslautern\\
	Kaiserslautern, Germany\\
	\texttt{akkoorath@cs.uni-kl.de}}
\affil[2]{Technical University of Kaiserslautern\\
	Kaiserslautern, Germany\\
	\texttt{bieniusa@cs.uni-kl.de}}
\affil[3]{Universidade do Minho\\
	Braga, Portugal\\
	\texttt{pg30467@alunos.uminho.pt}}
\affil[4]{HASLab, Universidade do Minho \& INESC TEC\\
	Braga, Portugal\\
	\texttt{cbm@di.uminho.pt}}

\usepackage{tikz}
\usetikzlibrary{arrows,decorations.pathmorphing,backgrounds,
	fit,positioning,shapes.symbols,shapes,shapes.geometric, chains,shapes.multipart, calc}
\usepackage{listings}
\usepackage{subfig}
\usepackage{amsmath}
\usepackage{amssymb}
\usepackage{xcolor}
\definecolor{light-gray}{gray}{0.95}

\newcommand{\seq}{{\ensuremath{\cdot}}}

\newcommand{\model}{global-local view}

\begin{document}

	\maketitle
	
	\begin{abstract}
		Concurrent linearizable access to shared objects can be prohibitively expensive in a high contention workload. Many applications apply ad-hoc techniques to eliminate the need of synchronous atomic updates, which may result in non-linearizable implementations.
		We propose a new programming model which leverages such patterns for concurrent access to objects in a shared memory system. In this model, each thread maintains different views on the shared object - a thread-local view and a global view. As the thread-local view is not shared, it can be updated without incurring synchronization costs. These local updates become visible to other threads only after the thread-local view is merged with the global view. This enables better performance at the expense of linearizability. We show that it is possible to maintain thread-local views and to perform merge efficiently for several data types and evaluate their performance and scalability compared to linearizable implementations. Further, we discuss the consistency semantics of the data types and the associated programming model.
		
	\end{abstract}

	
	\section{Introduction}
	
	Concurrent programming on shared-memory architectures is notoriously difficult.
	A concurrent system consists of a set of processes communicating implicitly through shared data structures.
	The visibility of updates on these data structures depends on the intricate interplay of synchronization mechanisms as defined by the memory model.
	Linearizability~\cite{Herlihy:1990:LCC:78969.78972} has turned out to be a fundamental notion on simplifying the reasoning about correctness of shared data structures for programmers.
	This consistency model formalizes the notion of atomicity for high-level operations.
	In an execution, every method call is associated with a linearization point, a point in time between its invocation and its response.
	The call appears to occur instantaneously at its linearization point, behaving as specified by the sequential definition.
	
	While linearizability is very useful for reasoning about the correctness of concurrent data structures, its implementation can be prohibitively expensive.
	As the number of cores increases in a multi-core system, the synchronization cost becomes more
	apparent that, it favors the relaxation of the concurrent objects semantics for scaling the programs \cite{Shavit:2011:DSM:1897852.1897873}.
	In practice, programming patterns are emerging that attempt to limit the associated cost of the required synchronization on the memory accesses.
	For example, in the widely-used messaging library ZeroMQ, adding messages to the queue is at the core of the application.
	While lock-free linearizable queues are fast, the developers observed that enqueuing new messages was affecting the overall performance, especially in high contention workloads \cite{aosazeromq}.
	However, only the relative order of messages from a single thread are relevant for the semantics of the message queue;
	it is not necessary to maintain a strict order of enqueue operations when two independent threads try to insert messages concurrently into the queue.
	To overcome the performance penalty, the developers re-engineered their message queue such that multiple messages are added as a batch, thus using only one single atomic operation.
	
	For another example, consider a shared counter that is concurrently updated by several threads.
	The final value of the counter must include all increments performed, but the order of increments is not relevant since all increments are commutative.
	If each increment executed by each thread is an atomic operation made visible to all other threads, it can become a bottleneck limiting the performance of the program~\cite{Boyd-Wickizer:2010:ALS:1924943.1924944}.
	In many cases, it is sufficient to execute the increment on some thread-local variable and to apply a combined update to the shared object.
	
	%
	%
	
	In this paper, we propose a new programming model for shared objects that leverages the different views of an object, the
	\model{} model. In this model, each thread has a local view of the object which is isolated from other
	threads. Threads update and read the local view. The local updates, though visible in a local view, are
	made visible on a global view only after an explicit two-way merge operation is performed. The other
	threads observe these changes once they synchronize, by merge, their local view with the global view. As the
	local view is non-shared, the local updates can be executed without requiring synchronization, thus enabling
	better performance, albeit at the expense of linearizability.
	
	In addition to the local operations, the model also provides synchronous operations on the global
	view. Consider, for example, a queue where the enqueues have been executed on the local view. To guarantee
	that the elements are dequeued only once, dequeues are executed atomically on the global view.
	We call the operations that perform only on local view, \emph{weak operations} and those on global view, \emph{strong operations}.
	Combining operations on the global and the local views, we can build data types with customizable semantics on the spectrum between sequential and purely mergeable data types.
	\emph{Mergeable data types} provide only weak and merge operations; \emph{hybrid mergeable data types} offer both weak and strong operations.
	An application that uses a hybrid mergeable data type may use weak updates when a non-linearizable
	access is sufficient and can switch to use only strong operations when stronger guarantees are
	required.
	
	In distributed systems, mergeable data types \cite{Shapiro:2011:CRD:2050613.2050642, Burckhardt:2014:RDT:2535838.2535848} are already widely in use.
	In this setting, each replica can be concurrently updated without requiring any synchronization and can later be merged with
	other replicas, while it is guaranteed that all nodes reach a convergent state once all updates have been delivered.
	To our knowledge, the applicability of such data types in a multi-core shared-memory setting has not been explored systematically, yet.
	In previous work, we have demonstrated how such types can be employed in Software Transactional Memory to prevent aborts  by resolving conflicts automatically \cite{Akkoorath2015}.
	In another work, Doppel \cite{Narula:2014:PRC:2685048.2685088}, an in-memory multi-core database, uses a per-core replica of objects and type-specific merge operations to parallelize conflicting transactions.
	
	\paragraph*{Contributions}
	This paper makes the following contributions:
	\begin{enumerate}
		\item We propose a new programming model, \model{}, for scalable multi-threaded applications that implements an adaptable trade-off between update visibility and synchronization cost (Section \ref{sec:model}).
		\item We provide a unified operational model of mergeable and hybrid data types and give a formal definition of their consistency semantics (Section \ref{sec:consistency}).
		\item We discuss the implementation of a mergeable counter, a hybrid counter, and a hybrid queue (Section \ref{sec:data types}) and compare their scalability with their linearizable counterparts in both low and
		high contention workloads (Section \ref{sec:evaluation}).
	\end{enumerate}
	
	In our preliminary work \cite{Akkoorath:2016:HCO:2911151.2911158}, we propose a mergeable counter and bag implementation.
	In contrast, this paper explores the concept of mergeability in depth by providing a formal model, specifications of further data types and an experimental evaluation.
	
	\section{Related Work}
	\label{sec:relatedworks}
	
	\emph{Programming models}: Maintaining per-thread replicas and performing updates on them has been considered by different programming models in the literature. In Concurrent Revisions \cite{Burckhardt:2010:CPR:1869459.1869515}, a forked thread's state is initially, a copy of its parent thread's state.
	The forked thread makes changes on its copy which is merged to the parent thread when it is joined back.
	During the join, conflicting updates are resolved using type-specific merge operations. The focus of this work is on fork-join model, where threads can communicate their state only when they join their parent. In contrast, we provide a generic model for the data types where a two-way merge and strong updates can share states among the threads at any point in the execution, thus enabling the applications to tune their use.
	
	Global Sequence Protocol (GSP) \cite{burckhardt_et_al:LIPIcs:2015:5238} is a model for replicated and distributed data systems.
	Similar to our model, GSP has a global state which is represented as a sequence of operations. Each client stores a prefix of this global sequence.
	The updates by client are first appended to the local sequence of pending operations and then broadcast to other replicas using a reliable total order broadcast protocol which enforces a single order on the global sequence.
	Since GSP addresses a distributed system's system model, with no bounds on message delays, there is much less control on replica divergence and liveness of the global sequence evolution. In contrast, here we address a shared-memory concurrent architecture that allows to reason about bounds on divergence and stronger progress guarantees on the evolution of shared state.
	
	Read-copy-update (RCU) \cite{Hart:2006:MLS:1898953.1898956} is a synchronization mechanism to allow processes to read a shared object while a concurrent modification is in progress. Similar to our model, multiple versions of the object are maintained so that readers observe consistent state while a modification is in progress. However, RCU is suited only for a single writer-multiple readers scenario. Read-log-update (RLU) \cite{Matveev:2015:RLS:2815400.2815406} is an improvement over RCU that allows concurrent writers. Unlike our model, concurrent writes are serializable which is achieved by serializing the writes or by fine-grained locking.
	
	\emph{Relaxed consistency models}: Many models attempt to relax the strict semantics of linearizability\cite{Herlihy:1990:LCC:78969.78972} to achieve better performance.
	Quasi linearizability \cite{Afek:2010:QRC:1940234.1940273} allows each operation to be linearized at a different point at some bounded distance from its strict linearization point.
	For example, a queue that dequeues in a random order, but never returns empty if the queue is not empty, is a quasi linearizable queue.
	Quasi linearizability, thus allows more parallelism by allowing flexible implementations.
	Our work is complimentary to this model, allowing a flexible combination of strong and weak updates to achieve different consistency semantics.
	Weak and medium future linearizability \cite{Kogan:2014:FSD:2611462.2611496} is applicable to the data types implemented using futures which allow flexible reordering of the operations.
	Others models, such as k-linearizability \cite{Aiyer2005} and quiescent consistency \cite{Viotti:2016:CND:2911992.2926965}, also define the correctness based on some sequential history, possible reordered, of the operations.
	
	\emph{Mergeable Data Types}: The idea of concurrent updates to the replicas of an object and merging them to a convergent state was formalized by Conflict free Replicated Data Types (CRDTs) \cite{Shapiro:2011:CRD:2050613.2050642}, which are now widely used in distributed replicated data systems.
	The properties of CRDTs, such as commutative operations and a semi-lattice structure, guarantee that concurrent updates can be safely executed on different replicas and later merged to get a consistent state on all replicas. The high network latency and possible reordering of messages in distributed system resulted in properties of CRDTs much different from what is required in a shared memory system.
	In this paper, we show implementations of mergeable data types that are tailored for shared memory concurrent programs.
	
	Even though no consolidated theory on mergeable data types exists in the shared memory ecosystem, there have been systems that use such types with restricted properties.
	Doppel \cite{Narula:2014:PRC:2685048.2685088} is a multi-core database that uses a mechanism called phase reconciliation to parallelize conflicting transactions. When a high contention workload is detected, Doppel switches to a split phase where the transaction updates per-core copy of the objects.
	At the end of the split phase, per-core copies are merged.
	Only operations that are commutative are executed in the split phase, thus guaranteeing serializability.

	\section{Programming Model}
	\label{sec:model}
	The system we consider is built upon a classical shared-memory architecture as supported by specifications such as the C++ or Java memory models.
	We assume that the system consists of a variable number of threads.
	Any thread can spawn new threads that may outlive their parent thread.
	The system distinguishes two types of memory: local memory is associated to a single thread and can only be accessed by this thread; shared memory can be accessed by any thread.
	Communication and coordination between the threads are done via shared-memory objects; we assume that there are no side channels.
	In particular, spawned threads do not inherit local objects from their parents.
	
	Each shared object $o$ has a global copy that is accessible by all threads that have a reference to it. In addition, each thread has its own local copy of $o$. A thread may update and read its own local copy, but it is not accessible by any other thread. The local updates are incorporated into the global copy when a $\mathsf{merge}$ operation is executed. Conflicting (non-commutative) updates from concurrent threads are resolved by a type-specific merge operation. In addition to the local updates and reads, the model also provides updates and read directly on the global copy. This gives flexibility for the data type semantics and the implementation of the underlying data structure.
	
	An operation $\mathsf{opKind}$ on an object in the \model{} model can be formalized as a function
	\[\mathsf{opKind}_{t}(m,g,s_t,l_t) = (r,g',s'_t,l'_t)\]
	where $m$ comprises the (optional) type-specific update($u$) or query($q$) method applied on the object, $g$ denotes the shared global object on which the operation is applied, and $t$ is a thread identifier that refers to the non-shared local version $(s_t,l_t)$ of the object.
	Here, $s_t$ denotes a local snapshot of the shared object state $g$ which gets updated upon synchronization, and $l_t$ refers to the local updates not yet incorporated in the shared global state $g$. The operation returns a tuple $(r,g',s'_t,l'_t)$ where $r$ is the return value of the  method $m$ and the other variables refer to the updated global $g'$ and local state $s'_t,l'_t$. State variables \--- $g,s_t,l_t$ \--- are each modeled as a sequence of updates, initially empty; a sequence $x$ can be concatenated with another sequence $y$ (or a single update), denoted by $x \cdot y$.
	
	Following are the basic operations in the \model{} model; these  are type-independent:
	
	\begin{itemize}
		\item \textsf{pull}
		updates the local object snapshot with the global object state; local operations are not changed.
		\[\mathsf{pull}_t(g,s_t,l_t) = (\bot,g,g,l_t)\]
		
		\item \textsf{weakRead}
		returns the result of a type-specific read-only operation $q$ on the state obtained by applying local updates on the local snapshot.
		\[\mathsf{weakRead}_t(\mathit{q},g,s_t,l_t) = (\mathit{q}(s_t \cdot l_t),g,s_t,l_t)\]
		\item \textsf{strongRead}
		returns the result of a type-specific read-only operation $q$ on the state obtained by applying local updates on global state. Neither the global state nor the local snapshot are changed.
		\[\mathsf{strongRead}_t(\mathit{q},g,s_t,l_t) = (\mathit{q}(g \cdot l_t),g,s_t,l_t)\]
		\item \textsf{weakUpdate}
		applies the update method $u$ on the local copy without any synchronization to the global state.
		\[\mathsf{weakUpdate}_t(\mathit{u},g,s_t,l_t) = (s_t \cdot l_t \cdot \mathit{u} ,g,s_t,l_t \cdot \mathit{u})\]
		\item \textsf{strongUpdate}
		applies the update method $u$ on the global state atomically. The previous weak updates that are batched in $l_t$ are not merged at this point.
		\[\mathsf{strongUpdate}_t(\mathit{u},g,s_t,l_t) = (g \cdot \mathit{u}, g \cdot \mathit{u}, s_t, l_t)\]
		\item \textsf{merge}
		incorporates the local updates to the global states and updates the local snapshot.
		\[\mathsf{merge}_t(g,s_t,l_t) = (\bot,g',g',\bot )\] where $g' = merge(g, (s_t,l_t))$ and $merge$ is type specific merge operation. In general, if the updates are commutative, $g' = g \seq l_t$. The data types can also specify a conflict resolving $merge$ operation, in case of non-commutative concurrent updates.
	\end{itemize}
	
	While $\mathsf{weakRead}$ and $\mathsf{weakUpdate}$ act exclusively on the local copy, $\mathsf{strongRead}$ and $\mathsf{strongUpdate}$ act on the global state.
	The combination of these two operations supports flexible optimizations on each individual data type. For example, a queue can guarantee that an element is dequeued only once by executing dequeues in $\mathsf{strongUpdate}$. At the same time, enqueues can be applied as $\mathsf{weakUpdate}$ and merged later for better performance. For an integer counter, we may want to enforce a weak limit on the maximum value, i.e. its value should not diverge arbitrarily from the defined maximum value. Such a counter can use a $\mathsf{strongRead}$ to check the global value to adapt the merge frequency or to switch to a fully synchronized version.
	
	
	\section{Data Types}
	\label{sec:data types}
	
	Each mergeable type defines a subset of the basic operations from the \model{} model, depending
	on the semantics needed. A purely mergeable counter defines only $\mathsf{weak}$ operations and
	$\mathsf{merge}$, while a hybrid mergeable counter also defines $\mathsf{strong}$ operations. In
	this section, we discuss the specification of several data types and their
	implementation.
	
	\subsection{Specification}
	
	Given a sequential counter with methods $\mathit{inc}$ (increments the counter by 1), $\mathit{value}$ (returns the current value), a purely mergeable counter implements following operations.
	
	\begin{itemize}
		\item $\mathsf{weakValue}_t()$ = $\mathsf{weakRead}_t(\mathit{value}, \_, s_t, l_t)$
		\item $\mathsf{weakInc}_t()$ = $\mathsf{weakUpdate}_t(\mathit{inc}, \_, \_, l_t)$
		\item $merge(g,(s_t, l_t))$ = $g \seq l_t$
	\end{itemize}
	
	The merge appends the local increments to the global sequence $g$, because the increments are commutative.
	A hybrid mergeable counter defines the following operations in addition to the above ones.
	The applications may choose \textsf{weak} or \textsf{strong} operations dynamically based on
	different criteria.
	\begin{itemize}
		\item $\mathsf{strongInc}_t()$ = $\mathsf{strongUpdate}_t(\mathit{inc}, g, \_, \_)$
		\item $\mathsf{strongValue}_t()$ = $\mathsf{strongRead}_t(\mathit{value}, g, \_, l_t)$
	\end{itemize}
	
	A sequential queue has operations $\mathit{enqueue}(e)$ and $\mathit{dequeue}$. A hybrid mergeable queue with mergeable enqueue and synchronized dequeue defines the following operations:
	\begin{itemize}
		\item $\mathsf{enqueue}_t(e)$ = $\mathsf{weakUpdate}_t(\mathit{enqueue(e)}, \_, \_, l_t) $
		\item $\mathsf{dequeue}_t()$ = $\mathsf{strongUpdate}_t(\mathit{dequeue}, g, \_, \_) $
		\item $merge(g,(s_t, l_t))$ = $g \seq l_t$
	\end{itemize}
	
	In the above semantics, if the global copy is empty, $dequeue$ returns null even if there are local
	enqueue operations by the same thread which have not been merged yet.
	We can allow dequeue to include local enqueue operations by defining
	
	
	$$\mathsf{dequeue}_t() =  \mathsf{strongUpdate}_t(\mathit{dequeue}, g', \_, \_) \textrm{ with } (\_,g'\_,\_) = \mathsf{merge}_t(g,s_t,l_t).$$
	
	In this way we can combine the operations to give different semantics.
	For example, a queue with weak enqueue and weak dequeue may be useful if redundant dequeue is not a
	problem for the application.
	A queue with both strong enqueue and strong dequeue behaves as a linearizable queue.
	
	
	A grow-only bag is a set that allows only $add$ operation, and allows duplicate elements. A purely mergeable bag implements $\mathsf{weakAdd}$ and $\mathsf{merge}$.
	
	\subsection{Implementation}
	
	The implementation of (hybrid) mergeable data types consists of two parts \--- an object variable for
	local view and another for global view. Local view and global view may or may not be of same type. A generic pattern for implementing a mergeable data type \textsf{MDT} is given by the following (object-oriented programming inspired) pseudocode:
\begin{lstlisting}
type MDT {
  ThreadLocal  T1 localView;
  T2 globalView;
  //weakUpdate or weakRead
  op1(param){
    localView.op1(param);
  }
  //strongupdate
  op2(param){
    atomic { globalView.op2(param); }
  }
  merge(){
    atomic {
       globalView.merge(localView);
       localView.reset(globalView);
    }
  }
\end{lstlisting}
	The types of localview and globalview (\textsf{T1,T2}) may or may not be of same type.
	Local views are thread-local instances as identified by \texttt{ThreadLocal}.
	A variable specified as \texttt{ThreadLocal} exists per thread in the thread's private storage.
	Many programming languages support some form of thread-local storage (TLS).
	A mergeable data type can also implement its own thread local storage by mapping thread ids to different instances of the object.
	
	\texttt{atomic} refers to any synchronization mechanism such as mutex or lock-free techniques such
	as compare and swap or transactional memory that atomically executes the code block with in.
	\textsf{op1, op2} refers to the methods implementing object's update or query operations. \textsf{reset} updates the local view to the global view.
	
	\textsf{weak} operations are executed on the local view. The \texttt{ThreadLocal} descriptor
	guarantees that each thread is accessing its own private view.
	For some data types, local views are isolated from each other and the global view, by maintaining a
	full copy of the object in each view. For large data structures, such as list or trees, maintaining
	a full copy is not efficient. Thus the local views may contain references to parts of the data
	structures that are shared by other local views or global view. In most cases, the shared parts are
	not directly updated by the weak updates, but only read. For example, a $lookUp$ on a list may
	first traverse the locally added items and then the shared parts of the list which are conceptually
	part of its local view. The mechanisms to make sure that an update on the global view does not change the local views, if it is updating the shared part, depends on the data type semantics and the underlying data structure being used.
	We show designs of a few data types where this can be done efficiently and correctly without copying the entire data structure.
	
	\paragraph*{Counter}
	The global view of a mergeable counter is an integer $g$. The local view consists of a pair of integers $(s,l)$.
	The weak increments are collected in the variable $l$ and added to $g$ during the merge. This design is inspired on \emph{sloppy counters}~\cite{Boyd-Wickizer:2010:ALS:1924943.1924944}, while using a local counter per thread instead of per core.
	The following pseudocode shows the implementation of a counter.
\begin{lstlisting}
type Counter: {int g, ThreadLocal int s, ThreadLocal int l}
weakInc() {
  l++;
}
strongInc(){
  atomic {g++}
}
int weakValue(){
  return s+l;
}
int strongValue(){
  return g+l;
}
merge(){
  atomic {g += l; s = g; l = 0;}
}
\end{lstlisting}
	
	It is easy to extend this implementation to allow decrements, explicit arguments for increments/decrements, and generalize to other commutative monoids.
	
	\paragraph*{Grow-only bag}
	
\begin{figure}[]
		\begin{subfloat}[Two threads with different local views.]{
				\begin{tikzpicture}[list/.style={rectangle split, rectangle split parts=2,
					draw, rectangle split horizontal}, >=stealth]
				
				\node[draw,inner sep=2pt, font=\tiny\sffamily] (H) {head};
				\node[list, right=6pt of H] (A) {};
				\node[list, right = 6pt of A] (B) {};
				\node[right = 6pt of B,draw,inner sep=2pt] (D) {};
				\draw (D.north east) -- (D.south west);
				\draw (D.north west) -- (D.south east);
				
				\draw[->] (H) -- (A);
				\draw[*->] let \p1 = (A.two), \p2 = (A.center) in (\x1,\y2) -- (B);
				\draw[*->] let \p1 = (B.two), \p2 = (B.center) in (\x1,\y2) -- (D);

				\node[list, above=6pt of H] (B1) {};
				\node[list, left = 6pt of B1] (A1) {};
				\node[draw,inner sep=2pt, font=\tiny\sffamily, left = 6pt of A1] (H1) {T1};
				\draw[->] (H1) -- (A1);
				\draw[*->] let \p1 = (A1.two), \p2 = (A1.center) in (\x1,\y2) -- (B1);
				\draw[*->] let \p1 = (B1.two), \p2 = (B1.center) in (\x1,\y2) -| (A.135);
				
				\node[list, below=6pt of H] (B2) {};
				\node[list, left = 6pt of B2] (C2) {};
				\node[list, left = 6pt of C2] (A2) {};
				\node[draw,inner sep=2pt, font=\tiny\sffamily, left = 6pt of A2] (H2) {T2};
				\draw[->] (H2) -- (A2);
				\draw[*->] let \p1 = (A2.two), \p2 = (A2.center) in (\x1,\y2) -- (C2);
				\draw[*->] let \p1 = (C2.two), \p2 = (C2.center) in (\x1,\y2) -- (B2);
				\draw[*->] let \p1 = (B2.two), \p2 = (B2.center) in (\x1,\y2) -| (A.230);
				
				\end{tikzpicture}
			}
		\end{subfloat}
		\begin{subfloat}[After T1's local view is merged.]{
				\begin{tikzpicture}[list/.style={rectangle split, rectangle split parts=2,
					draw, rectangle split horizontal}, >=stealth]
				
				\node[draw,color=white] (H) {};
				\node[list, right=6pt of H] (A) {};
				\node[list, right = 6pt of A] (B) {};
				\node[right = 6pt of B,draw,inner sep=2pt] (D) {};
				
				\draw (D.north east) -- (D.south west);
				\draw (D.north west) -- (D.south east);
				
				\draw[*->] let \p1 = (A.two), \p2 = (A.center) in (\x1,\y2) -- (B);
				\draw[*->] let \p1 = (B.two), \p2 = (B.center) in (\x1,\y2) -- (D);

				\node[list, above=6pt of H] (B1) {};
				\node[list, left = 6pt of B1] (A1) {};
				\node[draw,inner sep=2pt, font=\tiny\sffamily, left = 6pt of A1, dotted] (H1) {T1};
				\draw[->] (H1) -- (A1);
				\draw[*->] let \p1 = (A1.two), \p2 = (A1.center) in (\x1,\y2) -- (B1);
				\draw[*->] let \p1 = (B1.two), \p2 = (B1.center) in (\x1,\y2) -| (A.135);
				
				\node[list, below=6pt of H] (B2) {};
				\node[list, left = 6pt of B2] (C2) {};
				\node[list, left = 6pt of C2] (A2) {};
				\node[draw,inner sep=2pt, font=\tiny\sffamily, left = 6pt of A2] (H2) {T2};
				\draw[->] (H2) -- (A2);
				\draw[*->] let \p1 = (A2.two), \p2 = (A2.center) in (\x1,\y2) -- (C2);
				\draw[*->] let \p1 = (C2.two), \p2 = (C2.center) in (\x1,\y2) -- (B2);
				\draw[*->] let \p1 = (B2.two), \p2 = (B2.center) in (\x1,\y2) -| (A.230);
				
				\node[draw,inner sep=2pt, font=\tiny\sffamily, left = 2cm of A] (NH) {head};
				\draw[->,color=red] (NH) -| (A1.230);
				
				\end{tikzpicture}
			}
		\end{subfloat}
		\caption{Mergeable grow-only bag.}
		\label{fig:bag}
	\end{figure}
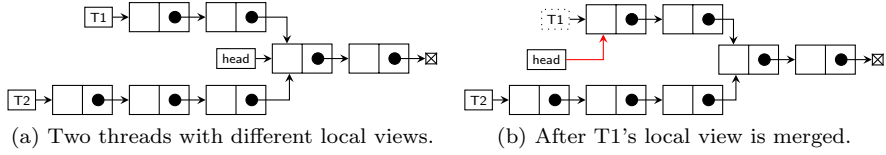

	A grow-only bag is implemented using a multi-headed list as shown in Figure \ref{fig:bag}. The thread local view consists of a pointer to the local head. A merge updates the global head of the list and does not change the local views of other threads. A lookup that traverses the list starting from the local head will never see an item that is concurrently added or merged.
	
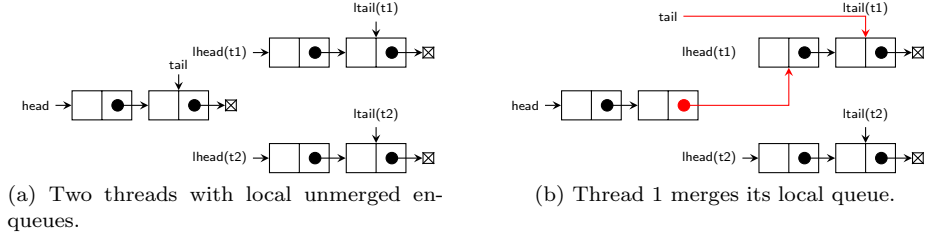
\begin{figure}[]
		\begin{subfloat}[Two threads with local unmerged enqueues.]{
				
				\begin{tikzpicture}[list/.style={rectangle split, rectangle split parts=2,
					draw, rectangle split horizontal}, >=stealth]
				
				\node[inner sep=2pt, font=\tiny\sffamily] (H) {head};
				\node[list, right=6pt of H] (A) {};
				\node[list, right = 6pt of A] (B) {};
				\node[right = 6pt of B,draw,inner sep=2pt] (D) {};
				\draw (D.north east) -- (D.south west);
				\draw (D.north west) -- (D.south east);
				\node[above = 6pt of B, inner sep=2pt, font=\tiny\sffamily] (L) {tail};
				
				\draw[->] (H) -- (A);
				\draw[*->] let \p1 = (A.two), \p2 = (A.center) in (\x1,\y2) -- (B);
				\draw[*->] let \p1 = (B.two), \p2 = (B.center) in (\x1,\y2) -- (D);
				\draw[->] (L) -- (B);

				\node[list, above right=17pt of D] (A1) {};
				\node[list, right = 6pt of A1] (B1) {};
				\node[inner sep=2pt, font=\tiny\sffamily, left = 6pt of A1] (H1) {lhead(t1)};
				\node[right = 6pt of B1,draw,inner sep=2pt] (D1) {};
				\draw (D1.north east) -- (D1.south west);
				\draw (D1.north west) -- (D1.south east);
				\draw[->] (H1) -- (A1);
				\draw[*->] let \p1 = (A1.two), \p2 = (A1.center) in (\x1,\y2) -- (B1);
				\draw[*->] let \p1 = (B1.two), \p2 = (B1.center) in (\x1,\y2) -- (D1);
				\node[above = 6pt of B1, inner sep=2pt, font=\tiny\sffamily] (L1) {ltail(t1)};
				\draw[->] (L1) -- (B1);
				
				\node[list, below right=17pt of D] (A2) {};
				\node[list, right = 6pt of A2] (B2) {};
				\node[inner sep=2pt, font=\tiny\sffamily, left = 6pt of A2] (H2) {lhead(t2)};
				\node[right = 6pt of B2,draw,inner sep=2pt] (D2) {};
				\draw (D2.north east) -- (D2.south west);
				\draw (D2.north west) -- (D2.south east);
				\draw[->] (H2) -- (A2);
				\draw[*->] let \p1 = (A2.two), \p2 = (A2.center) in (\x1,\y2) -- (B2);
				\draw[*->] let \p1 = (B2.two), \p2 = (B2.center) in (\x1,\y2) -- (D2);
				\node[above = 6pt of B2, inner sep=2pt, font=\tiny\sffamily] (L2) {ltail(t2)};
				\draw[->] (L2) -- (B2);
				
				\end{tikzpicture}
			}
		\end{subfloat}\qquad
		\begin{subfloat}[Thread 1 merges its local queue.]{
				\begin{tikzpicture}[list/.style={rectangle split, rectangle split parts=2,
					draw, rectangle split horizontal}, >=stealth]
				
				\node[inner sep=2pt, font=\tiny\sffamily] (H) {head};
				\node[list, right=6pt of H] (A) {};
				\node[list, right = 6pt of A] (B) {};
				\node[right = 6pt of B,inner sep=2pt] (D) {};
				\node[above = 24pt of B, inner sep=2pt, font=\tiny\sffamily] (L) {tail};
				
				\draw[->] (H) -- (A);
				\draw[*->] let \p1 = (A.two), \p2 = (A.center) in (\x1,\y2) -- (B);
				\draw[*->, red] let \p1 = (B.two), \p2 = (B.center) in (\x1,\y2) -| (A1);
				\draw[->, red] (L) -| (B1);

				\node[list, above right=17pt of D] (A1) {};
				\node[list, right = 6pt of A1] (B1) {};
				\node[inner sep=2pt, font=\tiny\sffamily, left = 6pt of A1] (H1) {lhead(t1)};
				\node[right = 6pt of B1,draw,inner sep=2pt] (D1) {};
				\draw (D1.north east) -- (D1.south west);
				\draw (D1.north west) -- (D1.south east);
				\draw[*->] let \p1 = (A1.two), \p2 = (A1.center) in (\x1,\y2) -- (B1);
				\draw[*->] let \p1 = (B1.two), \p2 = (B1.center) in (\x1,\y2) -- (D1);
				\node[above = 6pt of B1, inner sep=2pt, font=\tiny\sffamily] (L1) {ltail(t1)};
				
				\node[list, below right=17pt of D] (A2) {};
				\node[list, right = 6pt of A2] (B2) {};
				\node[inner sep=2pt, font=\tiny\sffamily, left = 6pt of A2] (H2) {lhead(t2)};
				\node[right = 6pt of B2,draw,inner sep=2pt] (D2) {};
				\draw (D2.north east) -- (D2.south west);
				\draw (D2.north west) -- (D2.south east);
				\draw[->] (H2) -- (A2);
				\draw[*->] let \p1 = (A2.two), \p2 = (A2.center) in (\x1,\y2) -- (B2);
				\draw[*->] let \p1 = (B2.two), \p2 = (B2.center) in (\x1,\y2) -- (D2);
				\node[above = 6pt of B2, inner sep=2pt, font=\tiny\sffamily] (L2) {ltail(t2)};
				\draw[->] (L2) -- (B2);
				
				\end{tikzpicture}
			}
		\end{subfloat}
		\caption{Hybrid mergeable Queue.}
		\label{fig:queue}
	\end{figure}

	\paragraph*{Queue}
	A hybrid mergeable queue can be implemented using a singly-linked list similar to a linearizable queue.
	The items enqueued are added to the tail of the list, while dequeue is performed from the head.
	A mergeable queue instance contains a global view \--- (\texttt{head, tail}), which points to the head and tail nodes respectively of
	the global list and local view \--- (\texttt{ThreadLocal lhead, ThreadLocal ltail}), which are the head and the tail
	of the local list of each thread.
	The local list collects the items enqueued by the thread that are not yet merged. The
	\textsf{merge}  atomically appends the local list to the global list (Figure \ref{fig:queue}).
	The time needed to merge a group of nodes is the same as the time needed to enqueue a single node.
	By batching the enqueues, we can reduce the number of synchronization operations, thus improving the overall throughput.
	
	The $dequeue$ operation directly updates the shared part of the list. For some data types, an update on the shared part of the data structure should preserve the old version, because local views may be keeping reference to it.
	However, there is no \textsf{weakRead}, such as a weak lookup, defined on queue that must observe a version before a concurrent dequeue. Hence, there is no need to keep those versions, which simplifies the implementation.

	\section{Correctness Definitions}
	\label{sec:consistency}
	
\newcommand{\defines}{\ensuremath{\triangleq}}
\newcommand{\MDTConsistency}{\textsc{GLConsistency}}

The data types designed using the \model{} model exhibit weaker consistency than linearizability. We
define the consistency model of mergeable and hybrid data types, named \MDTConsistency{}
, based on the notion of abstract executions, 
following the formalization in \cite{Viotti:2016:CND:2911992.2926965}.

An operation issued by a process on a shared object is represented by an event $e$, which is a tuple $(proc, kind, type, obj, ival, oval, stime, rtime)$, where
\begin{itemize}
	\item $proc$ is the id of the thread issuing the event.
	\item $kind$ denotes one of the operations defined in Section \ref{sec:model}. (\textsf{weakUpdate}, \textsf{weakRead} etc.).
	\item $type$ is the type-specific update or query method performed by the operation.
	\item $obj$ denotes the id of the object on which the operation is performed. 
	\item $ival$ refers to the input parameters for the update/query method.
	\item $oval$ is return value of update/query method.
	\item $stime$ is the event invocation time. We assume an abstract global time that can be used to determine relative ordering of events happening in concurrent. threads.
	\item $rtime$ is the event return time.
\end{itemize}

A history $H$ is a set of events. There are different relations defined on events in a
history. A session order $so$ is a partial order on the events. Two events $a,b$ are related
by $so$, $a \xrightarrow{so} b$, if both are invoked by the same thread and $a$ returns before $b$ is
invoked. For other relations, we refer to \cite{Viotti:2016:CND:2911992.2926965}.

An abstract execution is a multigraph  $A = (H, vis, ar)$. $vis$ is a partial order relation where $a
\xrightarrow{vis} b$ indicates that the effects of $a$ are visible to $b$. For example, if an increment
 operation is visible to a read, this means that the read returns a value of the counter obtained after
 executing the increment. $ar$ is a total order that
specifies how concurrent operations are ordered. For example, two concurrent merge operations may be
ordered based on the order of lock acquisition.

Further, the context of an event $cxt(A,e) \defines A|_{e, vis^{-1},vis,ar}$ encodes the events prior to $e$, which may influence its return value. The specification of a data type is given by a function $\mathcal{F}$ that determines the set of intended return values of an update or a query method in relation to its context.

We extend the formalism to specify \MDTConsistency as follows. $e.kind \in \{su, sr, wu, wr,$ $ pull, merge\}$ denotes the operations
\textsf{strongUpdate}, \textsf{strongRead}, \textsf{weakUpdate}, \textsf{weakRead}, \textsf{pull}
and \textsf{merge}. 
$ar|_{k}$ denotes the subset of $ar$ which involves only the operations where
$e.kinds \in k$ and $(a,b) \in ar|_{k_a \rightarrow k_b} \iff a.kind \in k_a
\land b.kind \in k_b \land a \xrightarrow{ar} b$. (Similarly, $so|_{k}$ defines the subset of $so$ restricted to $k$).
 
\MDTConsistency{} is defined per object. Henceforth, to simplify the notation, we assume that a history contains only operations on a single object. We can extend the definition to include a general history by extending predicates to restrict the operations on the same object. (For example, $ar|_{merge} \cap ob \subseteq vis$, where $a \xrightarrow{ob} b$ if $a,b$ are events applied on the same object). For a history $H$ and $\mathcal{A}$, the set of all abstract executions on $H$, we say that $H$ satisfies \MDTConsistency, if there exists an abstract execution $A \in \mathcal{A}$ such that $A$ satisfies the following predicate.
$$ \MDTConsistency(\mathcal{F}) \defines \textsc{GlobalOrder} \land \textsc{ThreadLocalOrder} \land \textsc{Vis} \land \textsc{RVal}(\mathcal{F})$$
$$ \textsc{GlobalOrder} \defines ar|_{su,merge} \subseteq vis \land ar |_{su,merge \rightarrow pull,sr} \subseteq vis$$
$$ \textsc{ThreadLocalOrder} \defines so|_{wu,wr,pull,merge} \subseteq vis \land so |_{wu \rightarrow sr} \subseteq vis$$
$$\textsc{Vis} \defines vis = ar|_{su,merge} \cup ar |_{su,merge \rightarrow pull,sr} \cup so|_{wu,wr,pull,merge} \cup so |_{wu \rightarrow sr}$$
$$ \textsc{RVal}(\mathcal{F}) \defines \forall op \in H : op.oval \in \mathcal{F}(op, cxt(A,op))$$

The updates on the global copy (\textsf{strongUpdate} and \textsf{merge}) are serializable. The
reads from the global copy (\textsf{strongRead }and \textsf{pull}) observe events in this order. This is
defined by \textsc{GlobalOrder}.
\textsc{ThreadLocalOrder} defines the visibility of thread local operations. The visibility of a thread's operation is defined by the session order
except for the \textsf{strongUpdate}, because \textsf{strongUpdate} is executing only on the global copy.
However, a \textsf{strongRead} observes the local \textsf{weakUpdate}s, which is captured by the predicate $so |_{wu \rightarrow sr} \subseteq vis$ .
In addition, the visibility relation is restricted by \textsc{Vis}, guaranteeing that two operations from
different threads are related by visibility only if there is a synchronous operation on the global
view between them. (Note that this predicate is the union of the predicates from \textsc{GlobalOrder}
and \textsc{ThreadLocalOrder}).

If thread A performs \textsf{weakInc;merge}, then thread B performs \textsf{pull;weakValue},
\textsc{GlobalOrder} guarantees that thread A's \textsf{merge} is visible to thread B's
\textsf{pull} which is again visible to its \textsf{weakValue} by \textsc{ThreadLocalOrder} and by
transitivity \textsf{weakInc} is visible to \textsf{weakValue}.
If the thread performs a \textsf{weakInc}, and then a \textsf{weakValue}, \textsf{weakInc} is visible to \textsf{weakValue}. However, a \textsf{strongInc} is not visible to a following \textsf{weakValue}, unless there is a \textsf{merge} or \textsf{pull} before the read.

\textsc{RVal} is the return value consistency, which guarantees that the return value of all operations belongs to intended values based on the specification $\mathcal{F}$.

	\section{Evaluation}
	\label{sec:evaluation}

	We evaluated the performance and scalability of the mergeable counter and the hybrid mergeable queue using different micro-benchmarks. As an example of real applications, we employed the hybrid queue in a breadth-first traversal on graphs. We implemented the counter in C++ and the queue in Java.
	
	The evaluations are performed on a 12 core 2.40GHz Intel(R) Xeon(R) CPU E5-2620 processor (2 NUMA nodes) with 2-way hyper-threading, under linux 4.4.0-62 Ubuntu x86\_64 and openjdk version 1.8.0\_121, clang version 4.0.0-svn297204-1, boost 1.58.0.1ubuntu1.
	
	\paragraph*{Counter}
	
\begin{figure}
		\includegraphics[width=\textwidth]{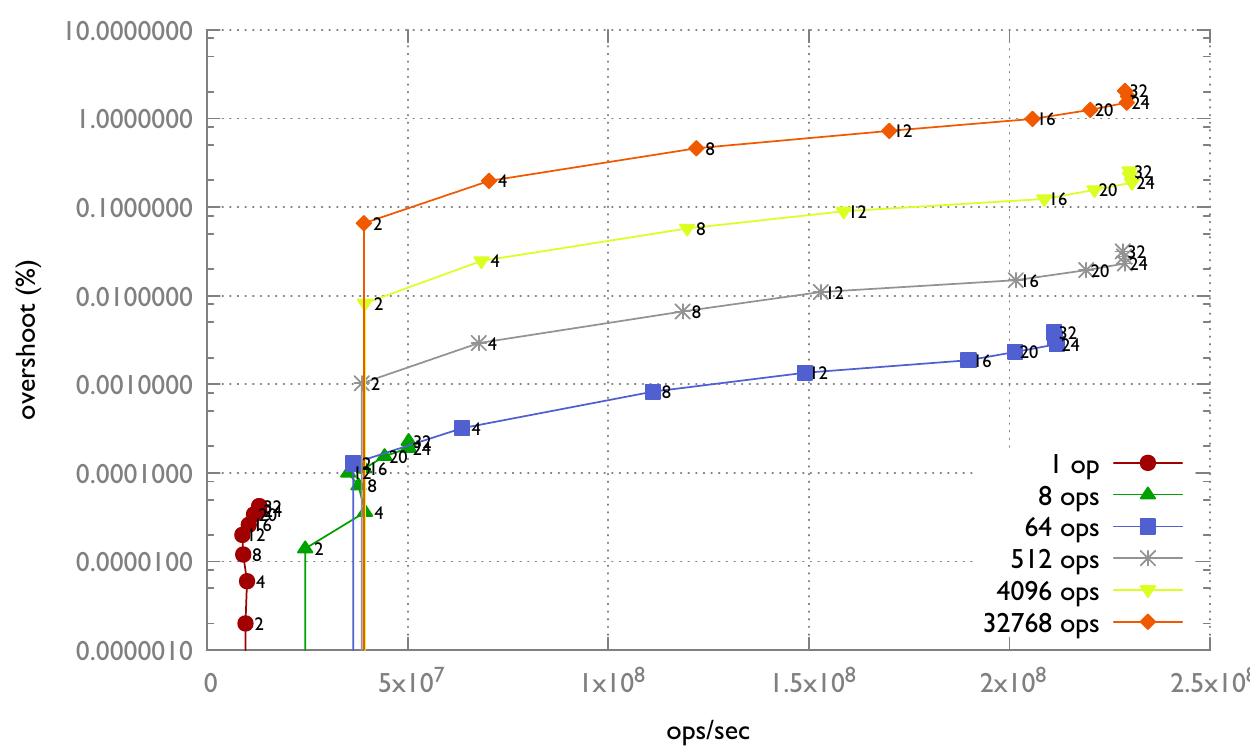}
		\caption{Throughput vs Overshoot of mergeable counter with different merge frequency.}
		\label{fig:counterovershoot}
	\end{figure}

	We provide two variants of a mergeable counter and compare them with an
	atomic counter, implemented using the atomic compare and swap operation. 
	The first
	version implements a weak mergeable counter, and is based on making threads
	increment their local count and periodically merge with the global count, also
	using atomics for efficiency.
	In the experiment, we allow threads to increment the shared mergeable counter
	until a \emph{target} value is reached. Since threads might not know about
	non-merged increments from other threads, they typically end up overshooting
	the target. For this experiment, the \emph{target} is set to $5 \times 10^6$ increments.
	We evaluated several merge
	frequencies, labeled with how many local increments are allowed between
	merges, and measured their throughput and the overshoot from the target.
	Figure \ref{fig:counterovershoot} shows that the throughput scales
	with the number of threads and with the merge frequency.  At the same time, the
	overshoot increases.  However, the percentage of the overshoot is small. 
	(Notice that overshoot is upper bound
	by the number of threads multiplied by the merge frequency, as this reflects at
	any given time the amount of increments not yet accounted for.)
	Points in the lines are labeled with the number of threads used. 
    As expected, the system does not scale beyond the point where the number of threads exceeds the number of cores (i.e at 24 threads).
    Also, note that for a single thread, overshoot is zero and thus the
	value is outside the logarithmic scale.
	
	\begin{figure}
		\begin{minipage}[b]{0.47\textwidth}
			\includegraphics[width=\textwidth]{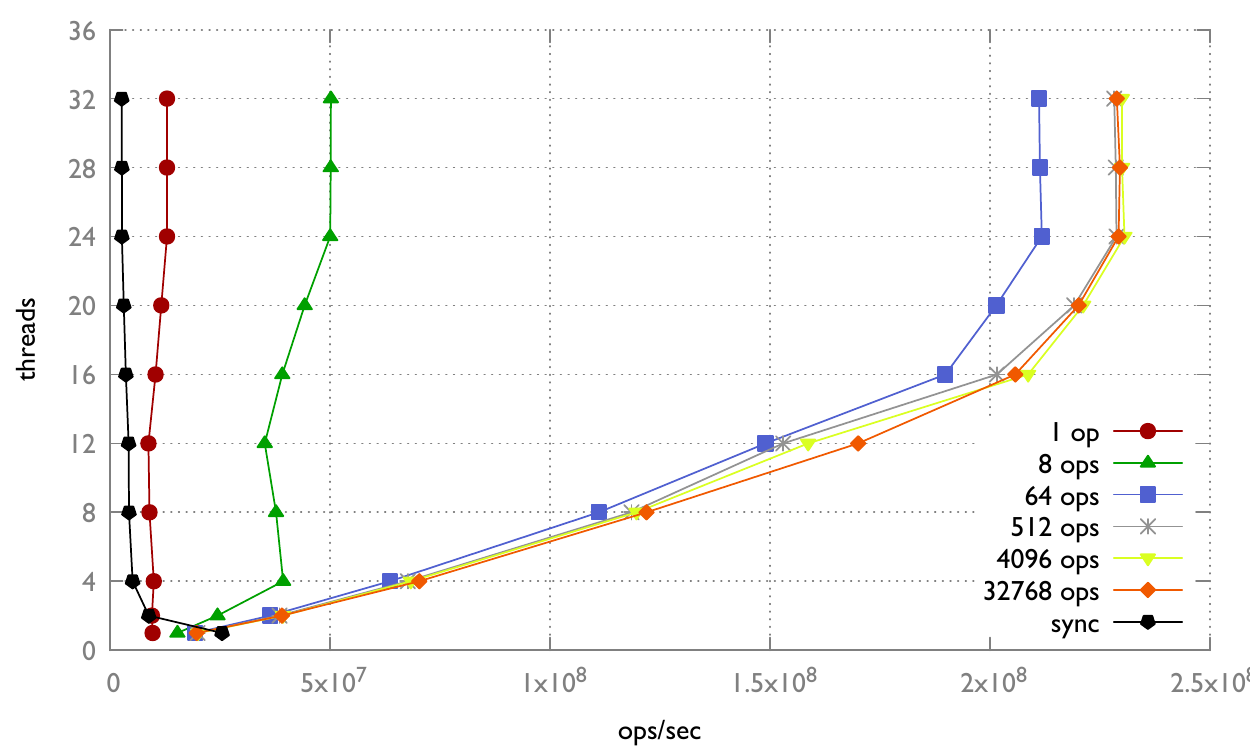}
			\caption{Throughput of mergeable counter vs atomic counter.}
			\label{fig:counteratomic}
		\end{minipage}\hfill
		\begin{minipage}[b]{0.47\textwidth}
			\includegraphics[width=\textwidth]{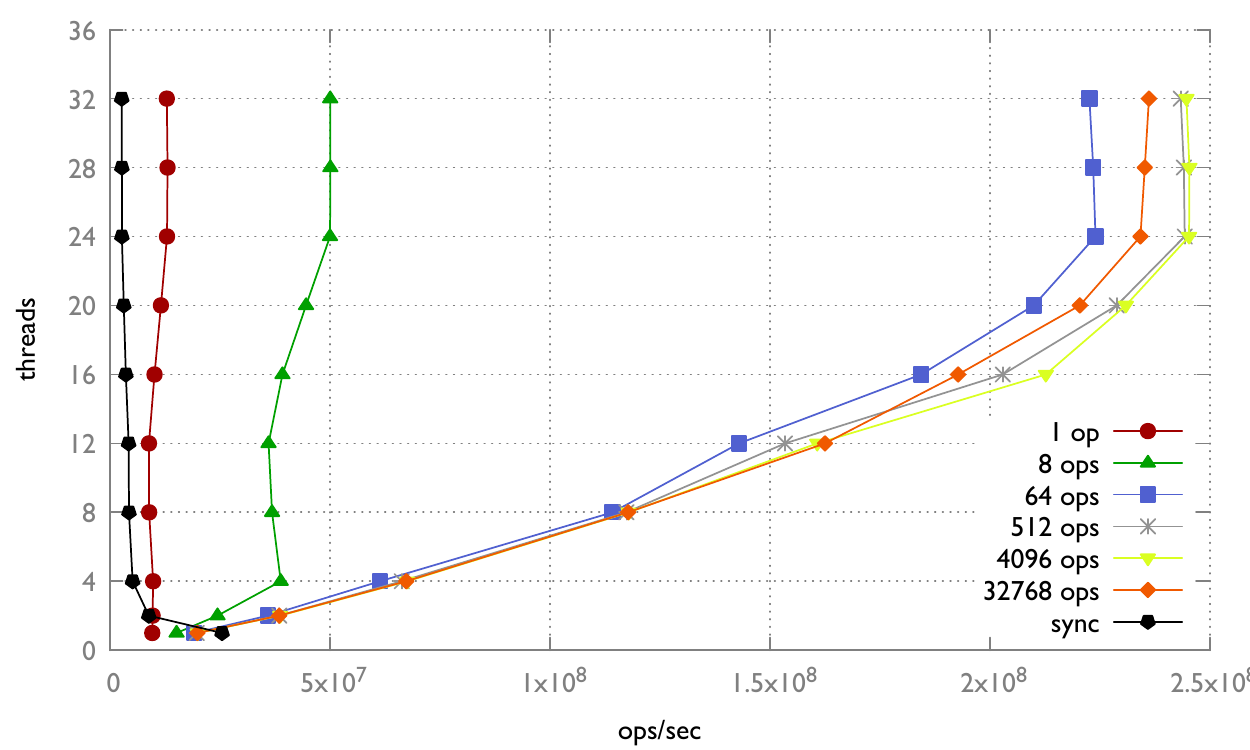}
			\caption{Throughput of hybrid mergeable counter (overshoot free) vs atomic counter.}
			\label{fig:counterhybrid}
		\end{minipage}
	\end{figure}
	
	Figure \ref{fig:counteratomic} shows the throughput of the mergeable counter compared to an atomic
	counter implemented using compare and swap. The atomic counter never overshoots the
	\emph{target}, but since threads are always competing on the increment, performance is very low and
	no speedup is obtained from multi-threading. In contrast, the mergeable counter can scale linearly
	up to a good fraction of the available concurrency, in particular with merge frequency of $\geq $ 4096. 
	
	While some applications could tolerate an overshoot, 
	in general, applications will require a
	tight target enforcement. To address this, we provide a variant of the mergeable counter that makes
	a hybrid use of initial weak local increments and later switches to atomic strong increments when
	approaching the target. 
	The first thread that, upon the periodic merges, detects that it is close to the target, initiates
	a barrier synchronization to ensure that all threads have switched to strong operations.
	Figure \ref{fig:counterhybrid} shows that under this approach, overshoot is eliminated while the
	performance is mostly identical to the mergeable counter.
	In general, the hybrid approach is efficient as long as the target is much larger
	than the merge frequency, since this limits the proportion of the execution done under
	linearizability. 
	
	

	
	\paragraph*{Queue}
	
\begin{figure}
		\begin{minipage}[b]{0.47\textwidth}
			\includegraphics[width=\textwidth]{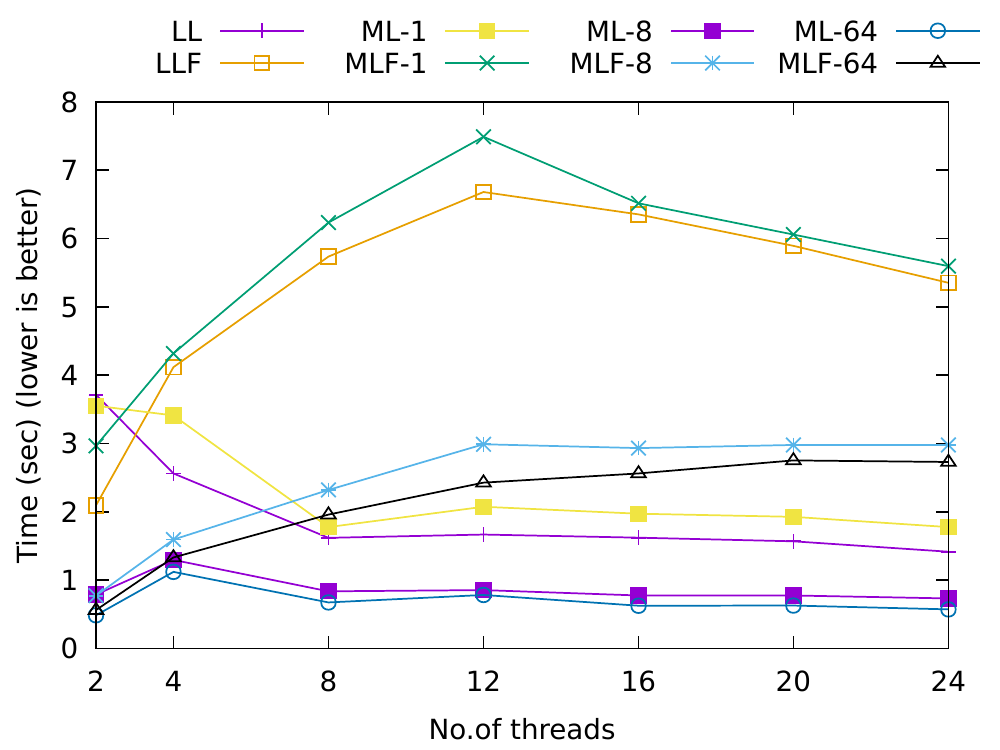}
			\caption{Queue. LL: linearizable lock-based, LLF: linearizable lock-free, ML: mergeable lock-based, MLF: mergeable lock-free. 1,8,64 - merge frequency for mergeable queues. \newline}
			\label{fig:queuebench}
		\end{minipage}	\hfill
		\begin{minipage}[b]{0.47\textwidth}
			\includegraphics[width=\textwidth]{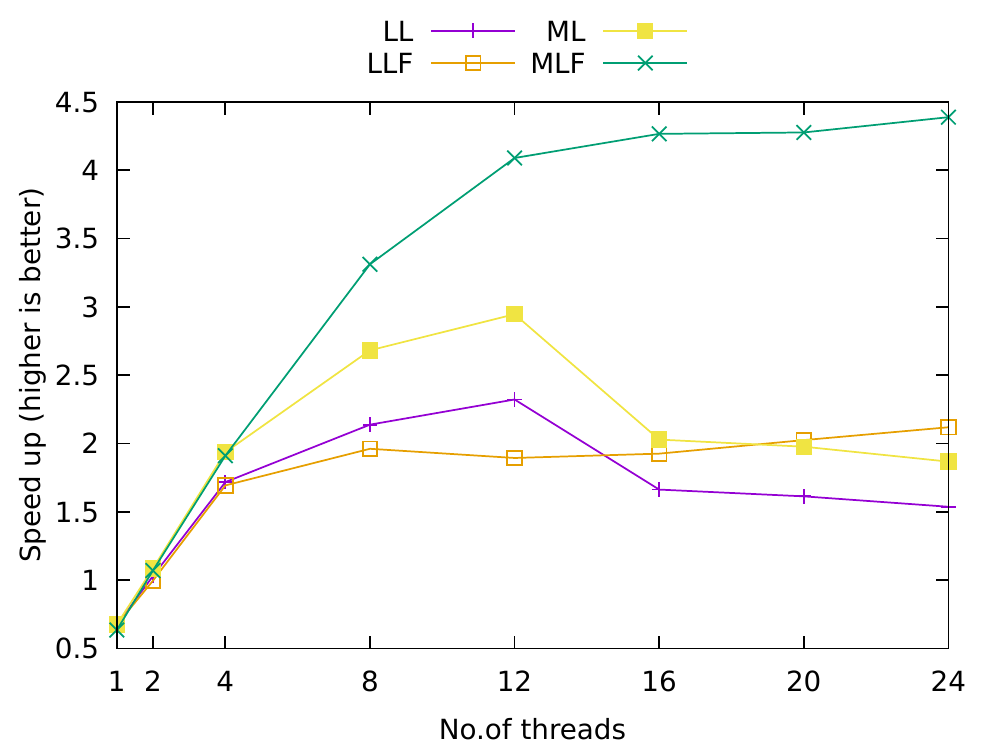}
			\caption{Breadth-first traversal on a graph using different queue implementations. LL: linearizable lock-based, LLF: linearizable lock-free, ML: mergeable lock-based, MLF: mergeable lock-free.}
			\label{fig:bfsbench}
		\end{minipage}
		
	\end{figure}

	To evaluate the scalability of hybrid mergeable queue (referred to as mergeable queue),
     we implemented four different queues in Java \--- 1) a lock-based linearizable queue based on Michael and Scott's 2-lock queue \cite{Michael:1996:SFP:248052.248106}, 2) a lock-based mergeable queue which uses similar 2-lock mechanism, 3) a lock-free linearizable queue adapted from Michael and Scott's lock-free queue \cite{Michael:1996:SFP:248052.248106} and 4) a lock-free mergeable queue.
	We evaluated the time to do a total of $5\times10^6$ enqueues and dequeues. Figure
	\ref{fig:queuebench} shows the result, evaluating mergeable queues with different merge
	frequencies $m$ (a merge is performed by a thread after $m$ enqueues). In this experiment, we forced half of the threads to run on one NUMA node and the other half on the second NUMA node. For both lock-based and
	lock-free versions, the mergeable queue is faster than the linearizable counterpart. Since this is a
	high-contention workload, the lock-based version performs better than the lock-free version. Unlike the mergeable
	counter, increasing merge frequency from 8 to 64 does not improve the performance significantly.
	The reason is that, $dequeue$ is always executed synchronously which shadows the performance gain from
	asynchronous $enqueue$s.
	
	\paragraph*{Breadth-First Traversal}
	A standard breadth-first traversal algorithm using queues can be parallelized using concurrent queues.
	We evaluated four versions of the algorithm using different queue implementations, that traversed randomly generated graphs of size of 2 $\times 10^6$ vertices and $2 \times 10^7$ edges. 
	Unlike the micro-benchmark for the queue, there is no fixed merge frequency. The threads merge their
	local queue at the end of processing each level.
	Figure \ref{fig:bfsbench} shows the speedup of each version compared to a single-threaded
	implementation. Mergeable queues scale better than their linearizable counterparts. 
	The speedup of the lock-free mergeable queue is significantly higher than that of the others, and scales almost linearly until 16 threads. Beyond 16 threads, the number of vertices processed by each thread at each level is reduced, as they are divided among the threads, leading to smaller merge frequencies. We believe the sudden drop in the speedup of lock-based queues after 12 threads is due to the additional cost in synchronization to the second NUMA core. Compared to the
	high-contention micro-benchmark from Figure \ref{fig:queuebench}, this is a low-contention workload because a significant amount of
	time is spent in processing the nodes rather than updating the queue.

	\section{Conclusion}
	
	An ever-increasing number of cores in combination with heterogeneous access latencies at different cache levels have advanced the spectrum of attainable performance from multi-thread programming. At the same time, this breaks the transparency with respect to data locality. As processor components become more numerous and spatially distributed, the cost of synchronization and communication among distant components will keep increasing in comparison to ones that are more closely located. When building internet-scale distributed systems, similar concerns lead to the design of scalable systems that limit global synchronization and operate locally when possible \cite{redblue180269}.
	
	Incorporating more information about the respective datatype semantics is crucial for datatype designs that are more parsimonious regarding synchronization. CRDTs succeed in capturing datatypes with clear concurrency semantics and are now common components in internet-scale systems. However, they do not migrate trivially to shared-memory architectures due to high computational costs from merge functions, which becomes apparent once network communication is removed.
	
	In this paper, we define the \emph{\model} model as base for a framework that allows capturing the semantics of multi-view datatypes.
	The \emph{\model} distinguishes between local fast state and distant shared state where operations need to be synchronized.
	This distinction allows the datatype designer to explore the trade-offs in the design when using weak or strong operations.
	Our approach enables speedups in order of magnitudes while preserving the datatypes' target behavior. We believe that the examples shown here are just the tip of the iceberg in terms of applicable datatypes.
	It is quite possible that further increments of the number of components involved will lead to a multi-tier model with more levels than the current binary, local vs global, scheme.

	\bibliography{references}
\end{document}